\documentclass{emulateapj}
\usepackage{apjfonts}
\journalinfo{The Astrophysical Journal, 2009, in press}

\shorttitle{PROTON \& ELECTRON HEATING IN THE SOLAR WIND}
\shortauthors{CRANMER ET AL.}

\begin{document}

\title{Empirical Constraints on Proton and Electron Heating in
the Fast Solar Wind}

\author{Steven R. Cranmer\altaffilmark{1},
William H. Matthaeus\altaffilmark{2},
Benjamin A. Breech\altaffilmark{3},
and
Justin C. Kasper\altaffilmark{1}}

\altaffiltext{1}{Harvard-Smithsonian Center for Astrophysics,
60 Garden Street, Cambridge, MA 02138}
\altaffiltext{2}{Department of Physics and Astronomy, Bartol
Research Institute, University of Delaware, Newark, DE 19716}
\altaffiltext{3}{NASA Goddard Space Flight Center, 8800 Greenbelt
Road, Greenbelt, MD 20771}

\begin{abstract}
We analyze measured proton and electron temperatures in the high-speed
solar wind in order to calculate the separate rates of heat deposition
for protons and electrons.
When comparing with other regions of the heliosphere, the fast solar
wind has the lowest density and the least frequent Coulomb collisions.
This makes the fast wind an optimal testing ground for studies of
collisionless kinetic processes associated with the dissipation of
plasma turbulence.
Data from the {\em Helios} and {\em Ulysses} plasma instruments were
collected to determine mean radial trends in the temperatures and
the electron heat conduction flux between 0.29 and 5.4 AU.
The derived heating rates apply specifically for these mean
plasma properties and not for the full range of measured values
around the mean.
We found that the protons receive about 60\% of the total
plasma heating in the inner heliosphere, and that this fraction
increases to approximately 80\% by the orbit of Jupiter.
A major factor affecting the uncertainty in this fraction is the
uncertainty in the measured radial gradient of the electron heat
conduction flux.
The empirically derived partitioning of heat between protons and
electrons is in rough agreement with theoretical predictions from
a model of linear Vlasov wave damping.  For a modeled
power spectrum consisting only of Alfv\'{e}nic fluctuations,
the best agreement was found for a distribution of wavenumber
vectors that evolves toward isotropy as distance increases.
\end{abstract}


\section{Introduction}

The supersonic solar wind is accelerated away from the Sun
by some combination of physical processes including gradients in
gas pressure (from the hot, $10^6$ K corona) and wave pressure,
as well as possible collisionless wave-particle interactions.
There is gradual heat input into the solar wind plasma that
begins in the corona and extends far out into interplanetary space
(see reviews by Parker 1963; Leer et al.\  1982; Tu \& Marsch 1995;
Goldstein et al.\  1995; Marsch 1999; Hollweg \& Isenberg 2002;
Cranmer 2002; Matthaeus et al.\  2003).
One likely source of this extended heating is the dissipation of
magnetohydrodynamic (MHD) turbulence.
In order to better understand the combined problem of coronal
heating, solar wind acceleration, and the large-scale evolution
of the turbulent heliospheric plasma, we need to know how energy
is transferred between the MHD fluctuations and the particles.

In the mainly collisionless solar wind, the various particle
species (i.e., protons, electrons, and heavy ions) are not in
thermal equilibrium with one another.
The particles exhibit a range of different outflow speeds,
temperatures, and velocity distribution anisotropies, and these
differences are most pronounced in low-density regions with
the least frequent Coulomb collisions (e.g., Neugebauer 1982;
Kohl et al.\  1997, 1998, 2006; Kasper et al.\  2008).
These differences can be used to probe the kinetic physical
processes that are responsible for depositing energy into
the plasma.

There have been a number of studies where the measured plasma
properties in the solar wind were used to derive the corresponding
rates of energy input from processes such as MHD turbulence
(e.g., Coleman 1968; Tu 1988; Freeman 1988; Verma et al.\  1995;
Matthaeus et al.\  1999b; Smith et al.\  2001; Vasquez et al.\  2007;
MacBride et al.\  2008; Marino et al.\  2008; Breech et al.\  2008).
Much of this work, however, dealt only with the energy budget of
the protons and not with any other ions or the electrons.
Although the solar wind's mass density and momentum flux are
dominated by the protons, the electrons carry approximately half
of the thermal energy of the plasma and should not be neglected in
a complete treatment.
There have been other investigations into the electron energy
balance in the solar wind (e.g., Scudder \& Olbert 1979;
Phillips \& Gosling 1990; Pilipp et al.\  1990).
There also have been studies of turbulent dissipation that compare
a total heating rate to the presumed proton contribution---and thus
estimate the electron heating rate as the remainder
(Leamon et al.\  1999; Stawarz et al.\  2009).
However, there has been surprisingly little {\em combined} analysis
of the proton and electron heating rates that treat the two plasma
components on equal footing.

In this paper we compute new estimates of proton and electron
heating rates in the fast solar wind.
These rates are derived from plasma properties measured by the
{\em Helios} and {\em Ulysses} spacecraft.
In {\S}~2 we describe these measurements in detail, and in {\S}~3
we show how the heating rates can be computed from the separate
equations of energy conservation for protons and electrons.
We find, not surprisingly, that an accurate determination of the
electron heating rate depends crucially on the measured electron
heat conduction flux.
In {\S}~4 we compare the empirically derived heating rates with
heating rates obtained from linear Vlasov theory, in which several
simple assumptions about the symmetry of an (Alfv\'{e}nic)
fluctuation spectrum were made.
Finally, {\S}~5 contains a summary of the major results of this
paper and a discussion of the implications these results may
have on our wider understanding of heliospheric plasma physics.

This work is being presented in tandem with an independent
investigation by Breech et al.\  (2009).
In both papers, we studied the same problem of proton-electron heat
partitioning in the fast wind, but it has been approached from
different and complementary vantage points.
This paper attempts to ``invert'' the in~situ measurements in order to
determine the partition fraction between proton and electron heating.
On the other hand, Breech et al.\  (2009) showed how a sophisticated
model of MHD turbulent heating---with an assumed value for the
partition fraction---can also be consistent with the empirical data.
The results of these two studies are consistent with one another.

\section{In Situ Particle Data}

We examine proton and electron plasma properties for the
high-speed solar wind between 0.29 and 5.4 AU.
Measurements made at larger distances (e.g., from the
{\em Voyager} probes) are excluded for two reasons:
(1) these data have been taken mainly in the ecliptic plane,
and thus are dominated by slow-speed wind, and
(2) the internal energy budget of particle velocity distributions
at distances $r \gtrsim 10$ AU is increasingly affected by pickup
ions, which are ignored here.
We focus on fast wind streams because these appear to be the
sites of the most ambient and ``quiescent'' solar wind plasma,
and because their low densities highlight them as the least
complicated by inter-species collisional coupling.

Figure 1 displays the data we use for the proton temperature $T_p$,
electron temperature $T_e$, and electron parallel heat conduction
flux $q_{\parallel, e}$.
Operationally, the high-speed wind was defined as being all
streams faster than 600 km s$^{-1}$.
This selection criterion was used for both the {\em Helios} and
{\em Ulysses} data sets.
A speed of 600 km s$^{-1}$ is slightly more restrictive than the
more standard value of 500 km s$^{-1}$ that has been applied by
others to define the fast wind (e.g., Dasso et al.\  2005;
MacBride et al.\  2008).
Taking a narrower range of speeds results in less contamination
from plasma parcels with qualitatively different properties.

\begin{figure}
\epsscale{1.17}
\plotone{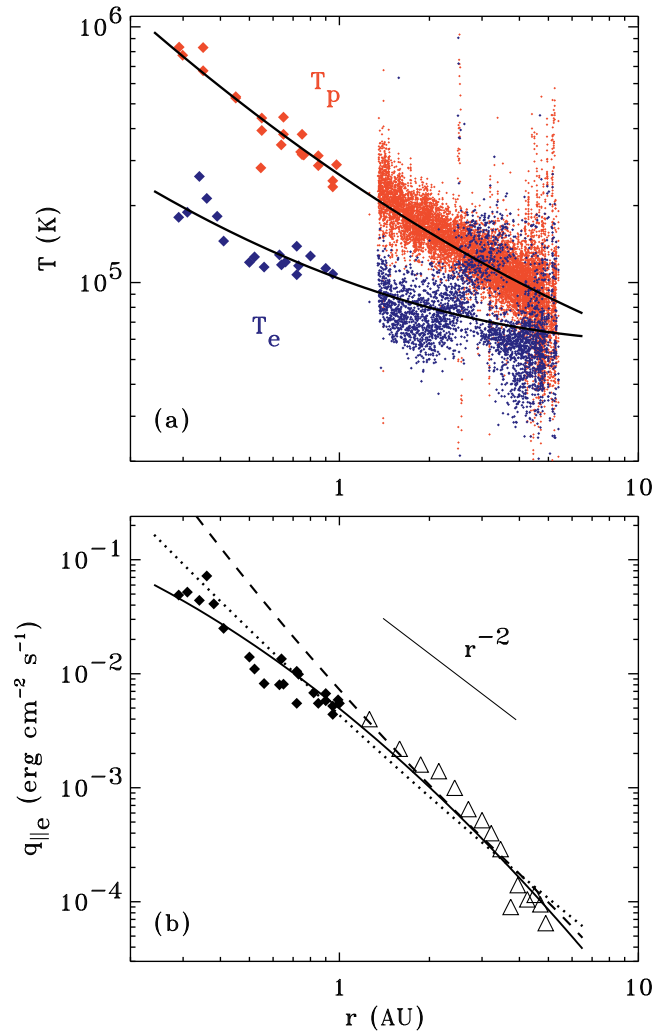}
\caption{In situ measurements for the fast solar wind:
({\em{a}}) plasma temperatures from {\em Helios}
({\em{filled diamonds}}) and {\em Ulysses} ({\em{small points}}),
with protons in red and electrons in blue.
({\em{b}}) electron heat conduction flux from {\em Helios}
({\em{filled diamonds}}) and {\em Ulysses} ({\em{triangles}}).
Also shown are least-squares fits ({\em{solid lines}}), the
classical Spitzer-H\"{a}rm heat flux ({\em{dashed line}}),
and the Hollweg (1974, 1976) collisionless heat flux with
$\alpha_{e} = 1.05$ ({\em{dotted line}}).}
\end{figure}

All measurements for heliocentric distances $r$ less than 1 AU came
from published data from the plasma instruments on the two
{\em Helios} spacecraft (Schwenn \& Marsch 1990).
Proton temperatures for the fast wind were obtained from
Marsch et al.\  (1982) and electron plasma properties ($T_e$
and $q_{\parallel, e}$) were taken from Pilipp et al.\  (1990).
The tabulated data points, which spanned the near-solar-minimum
years 1974--1976, corresponded to representative intervals that
were selected for broad coverage in distance and wind speed.
The actual velocity distributions for both protons and electrons
were non-Maxwellian, and the temperatures used here are mean
isotropic (drifting Maxwellian) equivalents, i.e.,
$T \equiv (T_{\parallel} + 2 T_{\perp})/3$.

Data for $r > 1$ AU came from the ``Solar Wind Observations Over
the Poles of the Sun'' (SWOOPS) instrument on {\em Ulysses}
(Bame et al.\  1992; Goldstein et al.\  1996).
The actual temperature measurements were extracted from the
European Space Agency's online archive.\footnote{%
The ESA Ulysses Data System (UDS) site for SWOOPS data is:
http://helio.esa.int/ulysses/archive/swoops.html}
We collected all SWOOPS data for the decade of time starting at
the beginning of the {\em Ulysses} mission (1990 November) and
extending to 2000 December.
This fully encompassed the solar minimum in 1996--1997, during
which there were found to be large swaths of the heliospheric
volume containing quiescent fast solar wind.
The highest time resolution data available for each interval was
used to better pinpoint regions of high-speed wind (using the
proton outflow speed $u_p$).
For the proton temperature data, this corresponded to a 4 or 8
minute cadence, depending on the mode of operation.
For the electron temperatures, the highest cadence tended to vary
between 7 and 35 minutes.

The {\em Ulysses} electron heat conduction data shown in
Figure 1{\em{b}} were taken from Figure 7 of Scime et al.\  (1994).
These measurements were taken in the ecliptic plane in 1990--1992
during the initial {\em Ulysses} cruise phase from 1 to 5 AU.
These data were not provided in a format where the high-speed
wind selection criterion ($u_{p} \geq 600$ km s$^{-1}$) could be
applied.
Thus, the points in Figure 1{\em{b}} at $r > 1$ AU apply to all
wind speed intervals.
However, Scime et al.\  (1999) found that there was no significant
variation in $q_{\parallel, e}$ as {\em Ulysses} passed between
fast and slow wind streams during its first high-latitude scan
in 1994--1995 (see also Salem et al.\  2003).

There is a drastic difference in the number of data points shown
in Figure 1{\em{a}} between the {\em Helios} and {\em Ulysses}
measurements.
For {\em Helios,} the high-speed velocity criterion resulted in
only 21 data points for $T_p$ and 18 data points for $T_e$.
For {\em Ulysses,} however, there were 211593 $T_p$ data points
and 61865 $T_e$ data points for the fast wind.
This is actually only a subset of the total amount of data present
in the archive, since we used only proton data validated by a
quality control flag that indicated nothing detectably wrong with
the measurement.
In order to keep Figure 1 from being unreasonably large in size,
we plotted every 20th individual SWOOPS temperature measurement.
For the collection of {\em Ulysses} data (chosen by the
criterion of $u_{p} \geq 600$ km s$^{-1}$), the mean wind speed
was found to be 744 km s$^{-1}$, with a standard deviation about
this mean of 47 km s$^{-1}$.
If the data selection criterion is reduced to $u_{p} \geq 500$
km s$^{-1}$, the mean speed and standard deviation change to
705 km s$^{-1}$ and 92 km s$^{-1}$, respectively.

There are a number of potential difficulties and systematic
uncertainties in measuring $T_p$ with the {\em Ulysses} plasma
instrument.
Some of these issues are more problematic for the slow wind,
since these regions tend to have lower values of $T_p$ and thus 
narrower distribution functions in velocity space.
These measurements may suffer from discretization effects
caused by the finite number of energy and angle channels of
the SWOOPS instrument.
However, like many solar wind instruments, SWOOPS has a
logarithmic distribution in the spacing of its energy channels.
This could result in an additional source of error for the fast
wind, since the velocity distributions are ``centered'' at higher
speeds where the relative energy resolution is coarser and
the protons are thus spread across fewer energy windows.

In order to convey a sense of the measurement uncertainties,
the SWOOPS data archive reports two independent determinations
of the proton temperature that are claimed to usually bracket
the true proton temperature from below ($T_{\rm small}$) and
from above ($T_{\rm large}$):
\begin{enumerate}
\item
The lower limit value $T_{\rm small}$ is essentially just the
radial component of the proton temperature, with the
high-energy ``beam'' component cut off.
Goldstein et al.\  (1996) concluded that the proton velocity
distribution is roughly isotropic---over the {\em Ulysses}
distances---because there was never a substantial dependence of
the radial temperature on the local magnetic field direction
(see also Vasquez et al.\  2007).
Thus, the closer the distribution tends to being isotropic in
velocity space, the better an assumption $T_{\rm small}$ will be.
\item
The upper limit value $T_{\rm large}$ was obtained by integrating
the measured proton velocity distribution over all of the energy
channels and angle bins that were statistically above a determined
noise level.
For the lowest temperatures, this process results in an
overestimate of $T_p$ because the angular responses of the SWOOPS
instrument channels may be broader than the actual velocity
distribution function.
\end{enumerate}

In Figure 1{\em{a}}, we adopt the geometric mean of these two
bounding values to obtain an intermediate estimate for
$T_{p} = (T_{\rm small} T_{\rm large})^{1/2}$.
It should be noted that there seems to be a solar cycle trend
in the ratio $T_{\rm large}/T_{\rm small}$.
At solar maximum (i.e., around 1990 and 2000), this ratio tends
to range between 2 and 3.
At solar minimum (i.e., 1996--1997), the ratio tends to range
between only 1.2 and 1.5.
Thus, the more we depend on the solar minimum SWOOPS data
(see below), the less we need to be concerned about the
differences between $T_{\rm large}$ and $T_{\rm small}$
(see also Marino et al.\  2008).

For the {\em Ulysses} electron data, we used the total temperatures
formed by a weighted sum of the narrow thermal core and the broader
``halo'' and ``strahl'' components of the velocity distributions.
The total electron temperature is the most consistent quantity
to use with the internal energy moment equations described
in {\S}~3.1, since these equations were derived by integrating
over all velocity space.
Figure 1{\em{a}} shows a noticeable bifurcation of $T_e$ into
hot and cold branches at distances around $r \approx 3$--4 AU.
This appears to be a solar cycle effect, since the lower values
are more dominant during 1996 (solar minimum) and the higher
values are more favored during the two solar maximum periods
(1991 and 2000) and, to a lesser extent, the post-maximum phase
(1994).
The higher $T_e$ data points at 3--4 AU also exhibited slightly
higher electron densities $n_e$ than the cooler data points.
Thus, the lower values seem more appropriate to be applied to
studies of the ambient fast solar wind associated with polar
coronal holes at solar minimum.
The least squares fit curve shown in Figure 1{\em{a}} for $T_e$
serendipitously favors these lower values.

Figure 1 also shows analytic fits for the measured quantities
between 0.29 and 5.4 AU.
These fits are given by
\begin{equation}
  \ln \left( \frac{T_p}{10^{5} \, \mbox{K}} \right) \, = \,
  0.9711 - 0.7988 x + 0.07062 x^{2}
  \label{eq:Tpfit}
\end{equation}
\begin{equation}
  \ln \left( \frac{T_e}{10^{5} \, \mbox{K}} \right) \, = \,
  0.03460 - 0.4333 x + 0.08383 x^{2}
  \label{eq:Tefit}
\end{equation}
\begin{equation}
  \ln \left( \frac{q_{\parallel, e}}{q_0} \right) \, = \, 
  -0.7032 - 2.115 x - 0.2545 x^{2}
  \label{eq:qefit}
\end{equation}
where $x \equiv \ln (r / [1 \, \mbox{AU}])$
and $q_{0} = 0.01$ erg cm$^{-2}$ s$^{-1}$.
To avoid the larger number of {\em Ulysses} data points
overwhelming the {\em Helios} data points in the least-squares
fitting process, the data sets from the two spacecraft were
weighted equally.
These fits should not be extended very far beyond the range of
heliocentric distances ($0.29 < r < 5.4$ AU) for which they were
derived.

The new fits given above can be analyzed in terms of a commonly
assumed power-law dependence of temperature as a function of
radius; i.e., $T \propto r^{-\delta}$.
The local value of the exponent $\delta$ can be computed at any
distance as the logarithmic derivative
$-\partial \ln T / \partial \ln r$.
Using equations (\ref{eq:Tpfit}) and (\ref{eq:Tefit}),
the resulting proton and electron temperature exponents $\delta_p$
and $\delta_e$ both decrease monotonically with increasing distance.
For protons, $\delta_p$ is approximately 0.98 at 0.3 AU, it
decreases to 0.80 at 1 AU, and then to 0.56 at 5 AU.
This range of values agrees with other exponents reported over
these distances that range between about 0.75 and 1 (e.g.,
Eyni \& Steinitz 1981; Lopez \& Freeman 1986;
Totten et al.\  1995; Ebert et al.\  2009).
For electrons, $\delta_e$ is approximately 0.65 at 0.3 AU, it
decreases to 0.43 at 1 AU, and to 0.15 at 5 AU.
These are also in agreement with earlier measurements of
order 0.2--0.6 (see Sittler \& Scudder 1980; Pilipp et al.\  1990;
Phillips et al.\  1995; Issautier et al.\  1998).
Some of these studies did not include any explicit criteria for
selecting either fast or slow solar wind streams, but it is
interesting that the same rough range of exponents is found.
It is possible to use these exponents to derive an effective
polytropic index for the protons and electrons---i.e.,
$\gamma = 1 + (\delta / 2)$ (Totten et al.\  1995)---but the
analysis in {\S}~3 attempts to do a more thorough study of
the internal energy balance of the plasma.

Figure 1{\em{b}} also shows how the measured electron heat
conduction flux compares to both the classical collisional heat
flux (Spitzer \& H\"{a}rm 1953) and to Hollweg's (1974, 1976)
collisionless ``free streaming'' approximation.
The former is given by
\begin{equation}
  q_{\parallel, e} \, = \, -\kappa_{e}
  \frac{\partial T_{e}}{\partial r}
\end{equation}
where the electron conductivity is
\begin{equation}
  \kappa_{e} = (1.84 \times 10^{-5} \,\, \mbox{erg}
  \,\, \mbox{cm}^{-1} \, \mbox{s}^{-1} \, \mbox{K}^{-7/2})
  \, \frac{T_{e}^{5/2}}{\ln \Lambda_{ee}}  \,\, .
\end{equation}
Also, the electron Coulomb logarithm is approximated by
\begin{equation}
  \ln \Lambda_{ee} \, = \, 23.2 + \frac{3}{2} \ln \left(
  \frac{T_e}{10^{6} \, \mbox{K}} \right) - \frac{1}{2} \ln \left(
  \frac{n_e}{10^{6} \, \mbox{cm}^{-3}} \right)
\end{equation}
where $n_e$ is the electron number density (see also
Cranmer et al.\  2007). 
The Spitzer-H\"{a}rm values are reasonably close to the measured
data above $r \approx 1$ AU, but they exceed the measurements at
smaller distances.
The Hollweg (1974, 1976) heat flux is given by
\begin{equation}
  q_{\parallel, e} \, \approx \, \frac{3}{2}
  \alpha_{e} n_{e} u k_{\rm B} T_{e}  \,\, ,
  \label{eq:alphac}
\end{equation}
where $k_{\rm B}$ is Boltzmann's constant.
The dimensionless order-unity constant $\alpha_e$ is only known
approximately, and it is expected to depend on the microscopic
shape of the electron velocity distribution (Hollweg 1974, 1976).
Using the above expression, though, we solved for $\alpha_e$ using
each of the data points in Figure 1{\em{b}}.
For simplicity, we used the above fit for $T_{e}(r)$, we also
assumed that $u = 700$ km s$^{-1}$, and we used the empirical
analytic model for $n_{e}(r)$ that is described below.
The mean value of $\alpha_e$ for all 38 points was 1.05, with
a standard deviation about the mean of 0.44 and no clear
radial trend.
(The full set of values ranged between 0.47 and 2.08.)
For comparison, Figure 1{\em{b}} shows a corresponding curve for
the collisionless heat conduction flux when $\alpha_{e} = 1.05$.

\section{Empirical Heating Rates}

\subsection{Internal Energy Conservation Equations}

Our goal is to use measurements to quantify the rates of heating
and cooling associated with as many physical processes as possible,
and then to solve for the net volumetric rates of heat {\em input}
that presumably can be attributed to MHD turbulence.
The primary mechanisms that are taken into account include
adiabatic energy conservation, electron heat conduction, and
Coulomb collisions.
We also retain the assumption that both the proton and electron
velocity distributions are isotropic Maxwell-Boltzmann distributions
with a shared bulk velocity ($u_{p} = u_{e}$).
Direct collisions between the proton and electron populations are
included for completeness, but they are expected to be extremely
weak in the heliosphere.
The models computed using classical collision rates (e.g.,
Spitzer \& H\"{a}rm 1953) are virtually identical to those
without any collisions whatsoever.
However, we also study the ramifications of more rapid collisions
by scaling up the classical rates by a range of constant factors.

It is important to discuss some of the physical processes that
are neglected in the models presented below.
For example, we do not explicitly consider how departures from
temperature isotropy (i.e., $T_{\parallel} \neq T_{\perp}$) affect
the energy balance.
Observationally, these departures are not significantly large over
most of the distances considered here (see
Marsch et al.\  1982; Pilipp et al.\  1990; Matteini et al.\  2007).
At 1 AU, Vasquez et al.\  (2007) showed that taking account of the
measured range of anisotropies would result in, at most, about a
10\% change in the adiabatic terms of the energy conservation
equations.
Still, it has been shown that for some regions in the solar wind,
the velocity distributions are kept from deforming too far away
from isotropy by plasma instabilities (e.g.,
Kasper et al.\  2002; Hellinger et al.\  2006).
We defer any consideration of the energy transfer between
instability-generated waves and the particles to future work.

Another process that is neglected here is proton heat conduction.
The classical proton heat conduction coefficient is about 25 times
smaller than that for electrons (Braginskii 1965;
Sandb{\ae}k \& Leer 1995).
Also, some models based on higher-order moment closures of the
Boltzmann equation have found even {\em lower} values than the
classical approach would suggest (e.g., Olsen \& Leer 1996).
The proton heat conduction flux was measured by {\em Helios} between
0.3 and 1 AU (Marsch et al.\  1982).
To confirm the expectation that the proton heat flux can be
neglected, we analyzed these measurements in terms of the
collisionless free-streaming approximation discussed above.
In other words, for each measurement we computed an effective
coefficient $\alpha_p$ analogous to the electron coefficient
$\alpha_e$ in equation (\ref{eq:alphac}).
Using the data tabulated by Marsch et al.\  (1982), we obtained
a mean value of $\alpha_{p} = 0.045$, which is a factor of
$\sim$23 lower than the mean electron coefficient $\alpha_{e} = 1.05$.
Test runs of the energy conservation models that included this
level of proton heat conduction resulted in heating rates that
were within 5\% of the ones computed without this effect.

We are now in a position to describe the conservation laws for
proton and electron internal energy upon which the present
results will be based.
Assuming a time-steady solar wind that has reached a constant
asymptotic terminal speed $u$ (see, e.g., Arya \& Freeman 1991),
these equations are
\begin{equation}
  \frac{3}{2} n_{p} u k_{\rm B} \frac{\partial T_p}{\partial r} -
  u k_{\rm B} T_{p} \frac{\partial n_p}{\partial r} = Q_{p} 
  + \frac{3}{2} n_{p} k_{\rm B} \nu_{pe} ( T_{e} - T_{p} )
  \label{eq:Ep}
\end{equation}
\begin{displaymath}
  \frac{3}{2} n_{e} u k_{\rm B} \frac{\partial T_e}{\partial r} -
  u k_{\rm B} T_{e} \frac{\partial n_e}{\partial r} =
  Q_{e} + \frac{3}{2} n_{e} k_{\rm B} \nu_{ep} ( T_{p} - T_{e} )
\end{displaymath}
\begin{equation}
  - \frac{1}{r^2} \frac{\partial}{\partial r}
  ( q_{\parallel, e} r^{2} \cos^{2} \Phi )
  \label{eq:Ee}
\end{equation}
where the heat input rates are $Q_p$ and $Q_e$, the Parker spiral
angle is $\Phi$, and the rates of proton-electron Coulomb
collisions for the two equations are $\nu_{pe}$ and $\nu_{ep}$
(see Barakat \& Schunk 1982; Isenberg 1984; Cranmer et al.\  1999).
As mentioned above, we assumed that the outflow speed $u$ is
constant and identical for the protons and electrons.
We also assumed that the electron densities $n_e$ and $n_p$ vary
with distance as $r^{-2}$, and that $n_p$ is normalized to
a value of 2.5 cm$^{-3}$ at 1 AU (e.g., Goldstein et al.\  1996).
We utilized a 5\% helium abundance by number in order
to compute $n_{e} = 1.1 n_{p}$.
Also, the winding angle of the spiral interplanetary magnetic
field is given in its standard form as
\begin{equation}
  \tan\Phi = \Omega r \sin\theta / u  \,\, ,
  \label{eq:Phi}
\end{equation}
where we used a rotation frequency $\Omega = 2.7 \times 10^{-6}$
rad s$^{-1}$.
In most of the models shown below, we set the colatitude
$\theta = 15\arcdeg$ to model the high-latitude {\em Ulysses}
measurements.\footnote{%
Although the {\em Helios} measurements were made close to the
ecliptic plane, the Parker spiral effect in the inner heliosphere
is not as pronounced as it is at $r \geq 1$ AU because of the
linear dependence on distance.  Thus, the resulting values of
$Q_e$ at the lowest values of $r$ are relatively insensitive to
the choice for $\theta$ (see Fig.\  4{\em{b}}).}

The Coulomb collision frequencies are balanced such that
$n_{e} \nu_{ep} \approx n_{p} \nu_{pe}$.
For two isotropic Maxwellian distributions interacting with one
another,
\begin{equation}
  n_{p} \nu_{pe} \, = \,
  \frac{32}{3} \pi^{1/2} \ln \Lambda \,
  \frac{e^{4} n_{e} n_p}{m_{e} m_{p} a^{3}}
\end{equation}
where the Coulomb logarithm $\ln\Lambda \approx 27$ at 1 AU,
$m_p$ and $m_e$ are the proton and electron masses, and
$e$ is the magnitude of the proton and electron charge
(Spitzer 1962).
The mean squared interaction speed is given by
\begin{equation}
  a^{2} \, = \, \frac{2 k_{\rm B} T_p}{m_p} +
                \frac{2 k_{\rm B} T_e}{m_e} \,\, .
\end{equation}
Note that, as defined above, the quantity $\nu_{pe}$ is a factor
of two larger than the equivalently named rate used by
Isenberg (1984) and Cranmer et al.\  (1999).
The present definition is more consistent with it being the true
rate of temperature equilibration as described by Spitzer (1962).
After evaluating many of the constants and using the approximation
that $m_{e} \ll m_{p}$, we found
\begin{equation}
  \nu_{pe} \, \approx \, 8.4 \times 10^{-9} \,
  \left( \frac{n_e}{2.5 \,\, \mbox{cm}^{-3}} \right)
  \left( \frac{T_e}{10^{5} \,\, \mbox{K}} \right)^{-3/2}
  \, \mbox{s}^{-1} \,\, .
  \label{eq:nupe}
\end{equation}
This expression gives rise to a rather large mean free path for
electron-proton collisions.
Depending on whether the mean free path is defined in terms of
the solar wind speed ($L_{\rm mfp} \sim u / \nu_{pe}$)
or the faster electron thermal speed
($L_{\rm mfp} \sim V_{th, e} / \nu_{pe}$), this quantity is
of order 500--1500 AU at $r = 1$ AU.
This should be contrasted with the much smaller mean free path
of $\sim$0.5 AU for electron-electron self-collisions that
maintain the thermal core at 1 AU (see, e.g., Spitzer 1962;
Salem et al.\  2003).

In addition to Coulomb collisions, there may be other collision-like
processes that lead to temperature equilibration and isotropization
in the heliosphere.
For example, collisionless wave-particle interactions have been
suggested for many years as being able to produce these effects
(Cuperman \& Harten 1970; Perkins 1973; Dum 1983;
Williams 1995; Kellogg 2000).
If the actual proton-electron temperature equilibration rate is
faster than expected, then {\em more} heat must go into the protons
for them to maintain the known inequality $T_{p} > T_{e}$ in the
fast wind.
Thus, if the collision rate is anomalously enhanced, the resulting
value of $Q_p$ would be larger (and $Q_e$ would be smaller) than
the values computed with weak or nonexistent collisions.
We include these effects below by multiplying the collision
rates $\nu_{pe}$ and $\nu_{ep}$ by an arbitrary constant $f$.

The electron heat conduction flux affects the energy balance in
qualitatively different ways depending on heliocentric distance.
Using the empirical fit for $q_{\parallel, e}(r)$
(eq.~[\ref{eq:qefit}]) and taking the divergence as shown in
equation (\ref{eq:Ee}), we found that this term leads to local
electron {\em cooling} for $r \lesssim 0.75$ AU, because the
radial slope of $q_{\parallel, e}$ is flatter than $r^{-2}$.
In this case, cool plasma at larger radii is conducted inward.
At larger distances, though, this term gives local electron heating,
since the slope of $q_{\parallel, e}$ is steeper than $r^{-2}$
and hot plasma is conducted outward.
Note, however, that if we had used either the classical
Spitzer-H\"{a}rm heat flux or the collisionless $\alpha_{e}=1.05$
approximation, as shown in Figure 1{\em{b}}, then
$q_{\parallel, e}$ would always be slightly steeper than
$r^{-2}$ and heat would be conducted outwards at all distances.

\subsection{Results for Proton and Electron Heating}

We solved equations (\ref{eq:Ep}) and (\ref{eq:Ee}) for the
volumetric heating rates $Q_p$ and $Q_e$ over the range of
heliocentric distances covered by the {\em Helios} and
{\em Ulysses} measurements.
An interesting aspect of this work is that because the internal
energy equations are not being solved for the temperatures---but
instead for the heating rates---we can avoid complicated
numerical differential equation techniques and use a
straightforward algebraic solution for $Q_p$ and $Q_e$.
Even the radial derivatives can be computed analytically from
the fits given in {\S}~2.
In practice, however, we computed the quantities to be
differentiated on a very fine grid (1000 points between 0.27
and 5.5 AU) and used a standard centered-difference approximation
to compute each of the terms in equations (\ref{eq:Ep}) and
(\ref{eq:Ee}).

The internal energy equations were first solved for two cases:
(1) standard Coulomb collision rates, with $f=1$, and
(2) a completely collisionless heliosphere, with $f=0$.
These two cases gave virtually identical results to one another,
with relative differences between the heating rates only at the
0.5\% level.
The procedure was also carried out several times for a range of
assumed (constant) wind speeds $u$ between 600 and 800 km s$^{-1}$.
One can see from equation (\ref{eq:Ep}) that, in the case of
negligible Coulomb collisions, our derived proton heating rate
$Q_p$ should be linearly proportional to the wind speed $u$.
The electron heating rate $Q_e$ also increases as the wind
speed increases, but because of the heat conduction this is
not a purely linear relationship.

Figure 2 shows the quantities $r^{4} Q_{p}(r)$ and $r^{4} Q_{e}(r)$
as a function of radius, rather than the rates themselves, because
the latter decrease very steeply with distance.
It is easier to see the subtle relative differences between $Q_p$
and $Q_e$ when the dominant radial variation has been removed.
It is clear that the computed values for $Q_{p}$ can be
well approximated by power law scaling relations in both
radius and wind speed.
The following fit was found to be valid to within about 6\%
relative accuracy over the full range of distances:
\begin{equation}
  Q_{p} \, \approx \, 3.42 \times 10^{-16}
  \left( \frac{r}{1 \,\, \mbox{AU}} \right)^{-3.5}
  \left( \frac{u}{700 \,\, \mbox{km s}^{-1}} \right)
  \,\, \mbox{erg} \,\, \mbox{s}^{-1} \, \mbox{cm}^{-3} \, .
\end{equation}
These scalings compare favorably to similar calculations by
Verma et al.\  (1995) and Vasquez et al.\  (2007).

\begin{figure}
\epsscale{1.17}
\plotone{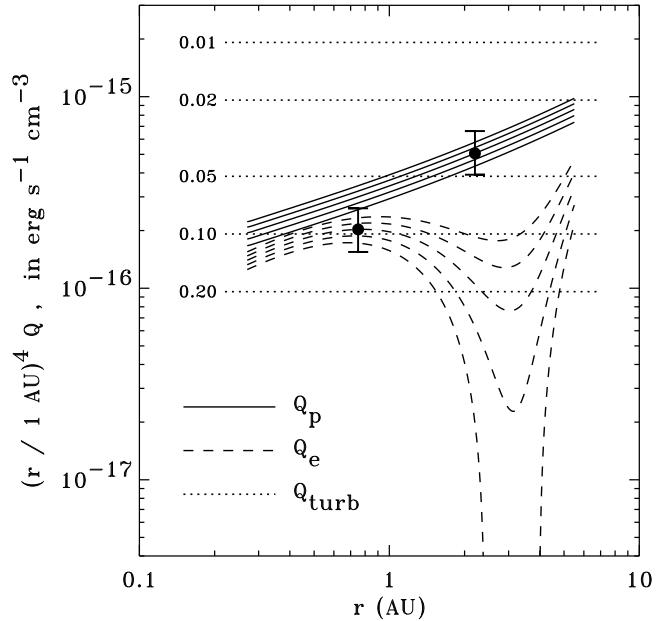}
\caption{Empirically derived heating rates for protons
({\em{solid lines}}) and electrons ({\em{dashed lines}}) in the fast
solar wind, with multiple curves showing results for $u = 600$,
650, 700, 750, and 800 km s$^{-1}$ (from bottom to top for each
set of curves).
Shown for comparison is $Q_{\rm turb}$ (eq.~[\ref{eq:Qturb}])
for five values of $\lambda_{\perp}$ at 1 AU ({\em{dotted lines}}).
All heating rates have been multiplied by $(r / 1 \, \mbox{AU})^4$.
Example error bars are given for the $u = 700$ km s$^{-1}$ case
(see text).}
\end{figure}

Note that it is not possible to fit the electron heating rate
$Q_{e}(r)$ with a simple power-law function because of the
nonlocal heat conduction.
The distances at which $Q_e$ decreases rapidly ($\sim$3 AU)
are roughly the same as where the ``bifurcation'' in $T_e$
occurs (see Fig.~1).
We cannot conclude whether this is a coincidence or not because
of the relatively sparse sampling of the fast wind by
{\em Ulysses} at these distances,
Note also that solving equation (\ref{eq:Ee}) gives negative
values for $Q_e$ at these heights for the smallest wind speed
of 600 km s$^{-1}$.
The relative impact of the electron heat flux is strong at these
distances, and the negative values would be eliminated if the
magnitude of $q_{\parallel, e}$ was smaller by only about 10\%.
This is certainly within reason, especially since the data used
to constrain the fit (Scime et al.\  1994) were not specifically
taken in the fast solar wind.

Figure 2 also includes representative uncertainty limits, shown
as error bars only at two distinct heights in order to distinguish
them clearly from the curves showing the variations in $u$.
These uncertainty limits reflect the existence of a range of
measured values around the mean fitting curves used for $T_p$,
$T_e$, and $q_{\parallel, e}$.
To compute these uncertainty bounds on $Q_p$ and $Q_e$,
we created a set of alternate models in which each of the
fits was was multiplied by factors $1.3^n$, where $n = -1, 0, +1$.
The result was a grid of 27 models in which the temperatures and
electron heat fluxes were varied up and down (in all combinations)
by fiducial $\pm$30\% amounts.
The extreme upper and lower limits on $Q_p$ and $Q_e$ were found
for all of these models and are plotted in Figure 2 at the two
example distances.
As can be seen from Figure 1, the actual spread in the in~situ
data often reaches---and sometimes exceeds---this fiducial 30\%
relative variability level.
Thus, it is clear that the results presented here about the
proton and electron heating rates depend crucially on the use of
the {\em mean radial trends} in the plasma parameters.
Further work is needed to ensure that these results are valid
for the totality of fast solar wind streams.
Other approaches to estimating the effect of variability of local
parameters on the heating rates include those of
Smith et al.\  (2006a) and Breech et al.\  (2008).

It is useful to compare the empirically derived heating rates
to those expected from a von K\'{a}rm\'{a}n similarity analysis
of the dissipation of an MHD turbulent cascade (e.g.,
Hossain et al.\  1995).
There are several pieces of evidence that suggest turbulence to
be responsible for the in~situ plasma heating (see, e.g.,
Coleman 1968).
The most direct evidence is the fact that $T_p$ in the solar
wind is positively correlated with the overall amplitude of
the turbulent fluctuations (Grappin et al.\  1990;
Vasquez et al.\  2007).
A more thorough discussion of the expected turbulent heating rate
is given by, e.g., Breech et al.\  (2008, 2009).
Roughly, though, we can estimate this rate as
\begin{equation}
  Q_{\rm turb} \, \approx \,
  \frac{\rho (Z_{+}^{2} Z_{-} + Z_{-}^{2} Z_{+})}{\lambda_{\perp}}
  \label{eq:Qturb}
\end{equation}
where the mass density $\rho \propto r^{-2}$ and the transverse
correlation length $\lambda_{\perp} \propto r^{1/2}$.
Note that the heating rate $Q_{\rm turb}$ depends on the cross
helicity of the fluctuations, which is defined by the ratio
$\sigma_{c} = (Z_{+}^{2} - Z_{-}^{2})/(Z_{+}^{2} + Z_{-}^{2})$,
and for highly Alfv\'{e}nic states where
$\sigma_{c} \rightarrow \pm 1$, we see that
$Q_{\rm turb} \rightarrow 0$.

The above phenomenological form for the turbulent heating rate has
been found to be consistent with numerical simulations of strong
MHD turbulence in a background magnetic field (e.g.,
Dobrowolny et al.\  1980; Hossain et al.\  1995;
Zhou \& Matthaeus 1990; Matthaeus et al.\  1999a;
Oughton et al.\  2001; Dmitruk et al.\  2001, 2002).
The success of transport theories in accounting for turbulence
properties observed by {\em Voyager,} {\em Ulysses,} and
{\em Pioneer} data (Zank et al.\  1996; Smith et al.\  2001;
Breech et al.\  2008) also provide empirical support for the use
of $Q_{\rm turb}$ in the form given by equation (\ref{eq:Qturb}).
The radial scaling $\lambda_{\perp} \propto r^{1/2}$ has been
measured in the heliosphere by, e.g., Bruno \& Dobrowolny (1986)
and Smith et al.\  (2001).
The measured Elsasser amplitudes $Z_{\pm}$ exhibit a complex radial
dependence that depends on various properties, but it is possible
to approximate them reasonably well with WKB Alfv\'{e}n wave
action conservation (Zank et al.\  1996).
Assuming spherical symmetry---which is appropriate for the
high-speed wind at high latitudes---this radial dependence is
$Z_{\pm} \propto r^{-1/2}$.
Putting these together we find that $Q_{\rm turb} \propto r^{-4}$,
which explains how the above approximations lead to the corresponding
curves in Figure 2 being constant (see also Dmitruk et al.\  2002;
Cranmer \& van Ballegooijen 2005).

To normalize $Q_{\rm turb}$ to the values shown in Figure 2, we
applied measured plasma properties at 1 AU, such as the proton
number density (2.5 cm$^{-3}$) and the Elsasser amplitudes
corresponding to outwardly propagating
($Z_{+} \approx 61$ km s$^{-1}$) and inwardly propagating
($Z_{-} \approx 26$ km s$^{-1}$) Alfv\'{e}n waves.
The latter two quantities combine to give a mean Elsasser
amplitude $Z \approx 47$ km s$^{-1}$ at 1 AU and a
normalized cross helicity $\sigma_{c} \approx 0.7$
(see, e.g., Bavassano et al.\  2000; Dasso et al.\  2005;
Cranmer \& van Ballegooijen 2005; Breech et al.\  2008, 2009).
The remaining free parameter in equation (\ref{eq:Qturb})
is the value of the transverse correlation length
at 1 AU, which we vary between 0.01 and 0.2 AU.
Figure 2 shows that the empirical heating rates for the fast wind
are consistent with $\lambda_{\perp} \approx 0.02$--0.1 AU.
Breech et al.\  (2008) found that this same range of correlation
lengths applies to the fast wind at 1 AU as well.\footnote{%
Multispacecraft measurements (e.g., Matthaeus et al.\  2005) show
the measured correlation scale at 1 AU to be somewhat smaller than 
the values in Figure 2.  This can be reconciled by using a value
of order 0.1--0.2 for the von K\'{a}rm\'{a}n proportionality
constant in equation (\ref{eq:Qturb}), instead of the value of
1 assumed here (see also Breech et al.\  2009).}

\begin{figure}
\epsscale{1.13}
\plotone{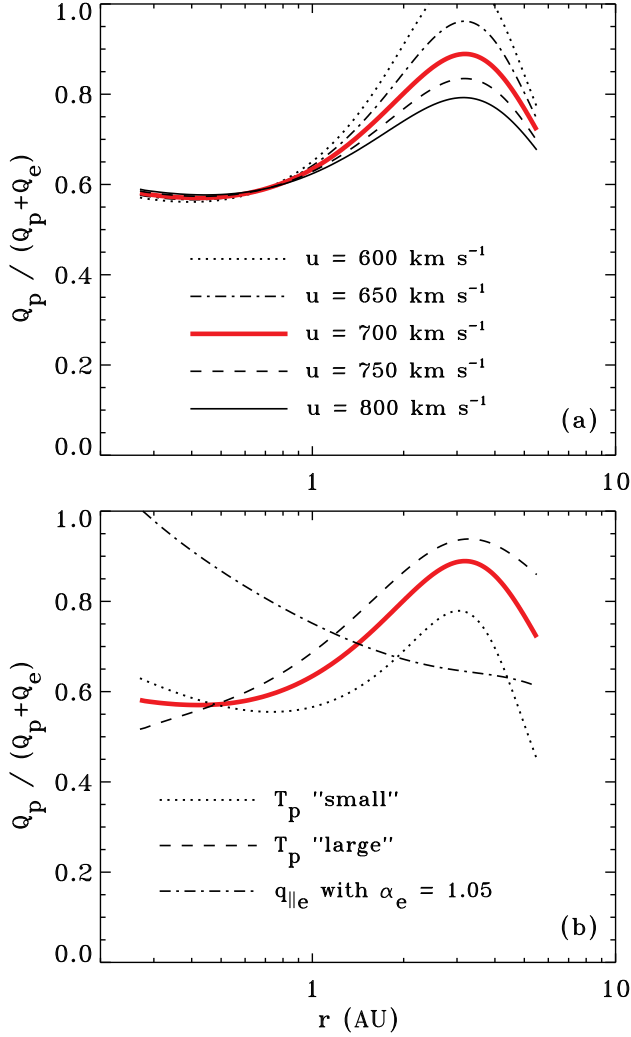}
\caption{Ratios of the proton heating rate to the total
(proton $+$ electron) heating rate.
The standard model with wind speed $u = 700$ km s$^{-1}$ is shown
in both panels ({\em{thick red line}}).
({\em{a}}) Variation of the wind speed between 600 and 800
km s$^{-1}$ in 50 km s$^{-1}$ increments (see labels).
({\em{b}}) Models computed with $T_{\rm small}$ ({\em{dotted line}})
and $T_{\rm large}$ ({\em{dashed line}}) for the SWOOPS proton
temperatures, and with equation (\ref{eq:alphac}) for
$q_{\parallel, e}$ instead of the empirical fit
({\em{dot-dashed line}}).}
\end{figure}

Figures 3 and 4 show the ratio of proton heating to the total,
i.e., $Q_{p} / (Q_{p} + Q_{e})$, for a number of different
calculations.
This fraction is denoted $f_p$ by Breech et al.\  (2009).
In all panels, a baseline model is shown for comparison
that was computed with $u = 700$ km s$^{-1}$ and the standard
choices for other parameters as described above.
Figure 3{\em{a}} shows the dependence of this ratio on the
assumed value of the wind speed.
The ratio at $r \lesssim 1$ AU appears to be insensitive to the
wind speed because both $Q_p$ and $Q_e$ vary linearly with $u$
at these distances (see also Fig.\  2).
The increased spread in the ratio at larger distances
is the result of the electron heat conduction having a larger
relative impact on $Q_e$ as the wind speed changes.

Figure 3{\em{b}} shows the dependence of the proton-to-total
heating ratio on varying several other assumptions of the
standard model (all keeping $u = 700$ km s$^{-1}$).
We computed trial versions of the proton temperature fitting curve
with the standard {\em Ulysses} measurements replaced by either
the lower or upper limit measurements ($T_{\rm small}$ or
$T_{\rm large}$ as described in {\S}~2).
We also tried replacing the fitting curve for the electron heat
conduction flux (eq.~[\ref{eq:qefit}]) with the simple
collisionless expression (eq.~[\ref{eq:alphac}]) and a constant
coefficient $\alpha_{e} = 1.05$.
This was seen in Figure 1{\em{b}} to possibly be a reasonable
description of the measured heat fluxes.

It is noteworthy that the use of the collisionless model for
$q_{\parallel, e}$ shows a rather extreme departure from the
other results, with the ratio $Q_{p} / (Q_{p} + Q_{e})$
monotonically decreasing as a function of increasing distance.
This difference arises because the collisionless heat flux
remains steeper than $r^{-2}$ in the inner heliosphere, and thus
its divergence (see the last term in eq.~[\ref{eq:Ee}]) gives
a positive contribution to the electron heating at all distances.
In a sense, the use of the collisionless heat flux is consistent
with the need for heat to be conducted outwards from heights
{\em below} our lower boundary at 0.3 AU.
The existence of this additional conductive heating below 1 AU
means that the deposited heating rate $Q_e$ need not be as large
as it would be otherwise, and thus the protons end up
contributing more to the total heating.
On the other hand, when using the least-squares fit to the measured
heat fluxes, the slope of $q_{\parallel, e}$ is flatter than
$r^{-2}$ in the inner heliosphere and conduction acts to cool
the electrons at these distances.
The derived value of $Q_e$ must then be larger, which gives the
rough equipartition between protons and electrons.
This is consistent with a more {\em local} deposition of heat to
the electrons that conducts both up and down from some point
around 0.75 AU.
We should emphasize, however, that the collisionless expression
is based on relatively simple theoretical scalings and it
is far from being a rigorous ``prediction'' for the electron
heat flux in the solar wind.
The discrepancy seen in Figure 3{\em{b}} suggests that the use
of a constant $\alpha_e$ coefficient in equation (\ref{eq:alphac})
should probably not be considered a robust approximation.

\begin{figure}
\epsscale{1.13}
\plotone{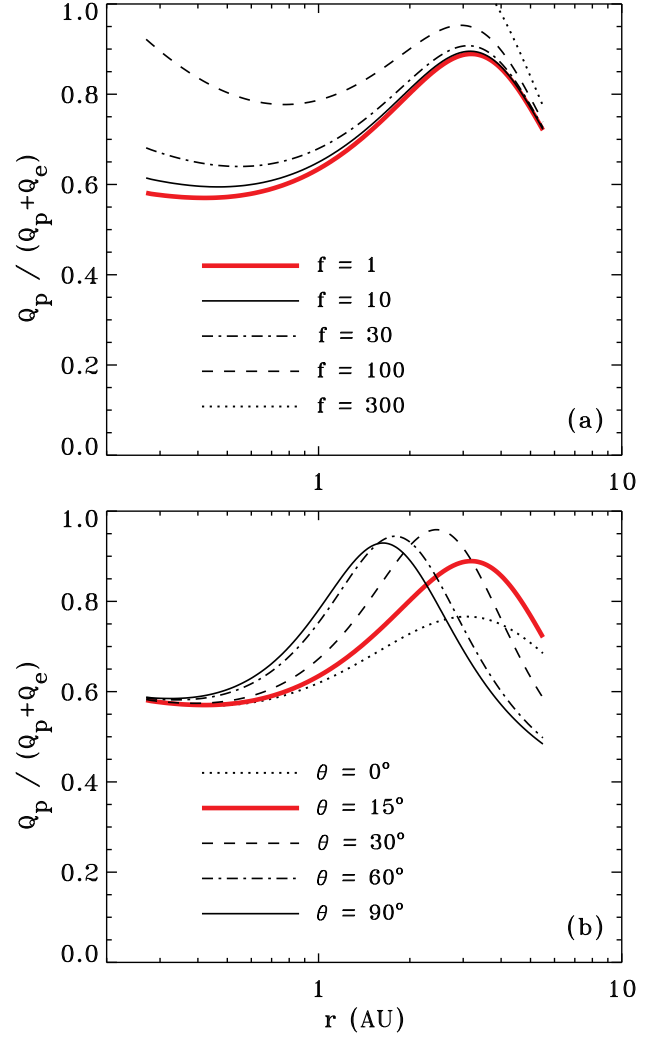}
\caption{Same as Figure 3, but with other parameters varied.
({\em{a}}) A series of models with anomalously strong Coulomb
collisions, with a range of constant multipliers $f$ to the
classical collision frequency (see labels).
({\em{b}}) Models computed over a range of colatitudes $\theta$
in the heliosphere, which affects the Parker spiral angle
$\Phi$ (see eq.~[\ref{eq:Phi}]).
The standard model with wind speed $u = 700$ km s$^{-1}$ is shown
in both panels ({\em{thick red line}}).}
\end{figure}

Figure 4{\em{a}} shows the result of varying the Coulomb collision
rate by multiplying $\nu_{pe}$ (eq.~[\ref{eq:nupe}]) by a range
of constant multipliers $f$.
As mentioned above, the curves for $f=0$ and $f=1$ would be
indistinguishable and the former is not shown.
As $f$ is increased, collisions attempt to equalize $T_p$ and
$T_e$ at a faster rate.
The fraction of heat going to the protons must increase in order
to maintain the specified $T_{p} >  T_{e}$.
Note, however, that above a certain value of $f$ there would
need to be net electron cooling, or essentially
$Q_{p} > (Q_{p} + Q_{e})$, in order to maintain the measured
temperature difference.
This allows us to rule out values of $f$ larger than about 200
over most heliocentric distances.
Figure 4{\em{b}} shows how the relative heating changes as the
assumed colatitude $\theta$ (and thus the Parker spiral angle
$\Phi$) is varied.
Models computed closer to the ecliptic plane exhibit a local
maximum in $Q_{p} / (Q_{p} + Q_{e})$ at smaller heliocentric
distances.

Given the uncertainties in many of the input parameters to these
models, the quantitative results regarding the relative
partitioning of $Q_p$ and $Q_e$ are also somewhat uncertain.
However, there does appear to be a preponderance of evidence
for the validity of two {\em qualitative} statements:
(1) In the inner heliosphere ($r < 1$ AU), there appears to be
rough equipartition between proton heating and electron heating.
(2) As heliocentric distance is increased from 0.3 to 5 AU, the
relative fraction of proton heating increases to noticeably
dominate the total heating rate.

\section{Expectations from Linear Alfv\'{e}n Wave Damping}

It is illustrative to compare the empirically derived partitioning
between proton and electron heating with theoretical predictions.
As a starting point for future work in this direction, we computed
some extremely simple estimates for the wavenumber dependence of a
turbulent power spectrum of Alfv\'{e}nic fluctuations.
These were then coupled to a general Alfv\'{e}n wave dispersion
relation that allowed us to compute the contributions to proton and
electron heating.
Although it is well-known that strong MHD turbulence is far from
``wavelike'' (i.e., one might expect that a coherent wave survives
for only about one period before nonlinear processes transfer its
energy to smaller scales), there is a long history of using damped
linear wave theory to study the small-scale dissipation of a
cascade (see, e.g., Eichler 1979; Quataert 1998;
Leamon et al.\  1999; Quataert \& Gruzinov 1999; Marsch \& Tu 2001;
Cranmer \& van Ballegooijen 2003; Gary \& Borovsky 2004, 2008).
A typical justification of this approach is that the amplitudes
of magnetic fluctuations in the dissipation range tend to be
extremely small ($\langle \delta B^{2} \rangle \ll B_{0}^2$);
see also Spangler (1991) and Miller et al.\  (1996).

We assumed that the wavenumber dependence of the power spectrum
of magnetic fluctuations $P_B$ scales as $k^{-7/2}$, where
$k$ is the magnitude of the wavenumber.
The power $P_{B}(k)$ is defined such that its integral over the
full (three-dimensional) wavenumber space gives the total magnetic
energy density of fluctuations.
The exponent of $-7/2$ corresponds to an exponent of $-3/2$ for
a corresponding one-dimensional isotropic spectrum (i.e.,
the latter measuring power in spherical shells in $k$-space).
This exponent has been proposed for various theories of MHD
turbulence (e.g., Iroshnikov 1964; Kraichnan 1965; Nakayama 2001).
This value is also close to the exponent of $-10/3$ that has
been predicted for a purely perpendicular cascade having a
one-dimensional power spectrum of $k_{\perp}^{-5/3}$
(see Kolmogorov 1941; Fyfe et al.\  1977; Higdon 1984;
Goldreich \& Sridhar 1995; Cho \& Vishniac 2000;
Boldyrev 2005; Horbury et al.\  2008).

We also assumed that the above wavenumber dependence applies in
restricted ranges of $\theta_{kB}$, which is defined as the angle
between the wavenumber vector and the background magnetic field
direction.
Specifically, we constructed three models with different angular
distributions of wave power:
(1) a ``slab'' spectrum, with nearly parallel-propagating waves
filling the region $0 \leq \theta_{kB} \leq {5\arcdeg}$,
(2) a ``two-dimensional'' (2D) spectrum, with nearly
perpendicularly-propagating waves filling the region
${85\arcdeg} \leq \theta_{kB} \leq {90\arcdeg}$, and
(3) an isotropic spectrum with all values of $\theta_{kB}$
having the same wavenumber dependence.

The second case given above---i.e., a turbulent cascade mainly
in the $k_{\perp}$ direction---uses a range of angles that is
actually quite {\em broad} in comparison to theoretical
expectations of the so-called ``critical balance'' proposed by
Goldreich \& Sridhar (1995).
This condition is defined by an equivalence between the time scales
of Alfv\'{e}n wave propagation along the field and nonlinear energy
transfer perpendicular to the field.
According to Goldreich \& Sridhar (1995), the majority of the
power in strong MHD turbulence should have angles between the
critical balance angle $\theta_{\rm crit}$ and {90\arcdeg}.
At large perpendicular wavenumbers $k_{\perp}$ (i.e., at the onset of
kinetic Alfv\'{e}n wave dispersion and substantial Landau damping),
we estimated that $\theta_{\rm crit}$ decreases slightly from
approximately {89.8\arcdeg} at 0.3 AU to {88.9\arcdeg} at 5 AU.
These values were computed from the existing models of
Cranmer \& van Ballegooijen (2003, 2005) and the measured
plasma parameters given in {\S}~2.

We computed the proton and electron heating rates for each kind
of spectrum using the quasi-linear framework developed by
Quataert (1998), Quataert \& Gruzinov (1999), Marsch \& Tu (2001),
and Cranmer \& van Ballegooijen (2003).
The heating rates are given by integrals over vector wavenumber
$\bf k$ of the form
\begin{equation}
  Q_{s} \, = \, \rho \int d^{3} {\bf{k}} \, P_{\rm tot}({\bf{k}})
  \, 2 \omega_{i,s}
\end{equation}
where $s = p,e$ denotes the particle type of interest and
$P_{\rm tot} \approx 2 P_{B}$ is the total energy spectrum of
fluctuations.
The species-dependent amplitude damping rates $\omega_{i,s}$ were
defined by Cranmer \& van Ballegooijen (2003) to be weighted by
Landau/cyclotron resonance functions that take account of the
particle kinetic motions.
For a proton-electron plasma, the sum $\omega_{i,p} + \omega_{i,e}$
gives the absolute value of the total linear damping rate (i.e.,
the imaginary part of the wave frequency).
The resonance functions depend on the dispersive properties of
the Alfv\'{e}n waves, which were computed using the warm-plasma
Vlasov-Maxwell dispersion code of Cranmer \& van Ballegooijen (2003).
These functions also depend on the shape of the particle velocity
distributions, which were assumed here to be Maxwellian.

In order to compute the {\em relative} heat dissipated by the
protons versus that dissipated by the electrons, we did not need
to specify an absolute normalization for the power spectrum.
However, in practice, it was useful to cut off the spectrum at
low wavenumbers using an ``outer scale'' lower limit $k_{\rm out}$
that scaled inversely with the turbulent correlation length
discussed above; i.e.,
$k_{\rm out} \propto \lambda_{\perp}^{-1} \propto r^{-1/2}$.
The upper limits for the parallel and perpendicular wavenumbers
($k_{\parallel}, k_{\perp}$) were found numerically by the
Vlasov-Maxwell code as locations where the Alfv\'{e}n-wave
solution branches ceased to exist (see Stix 1992;
Cranmer \& van Ballegooijen 2003).

The dispersive properties of Alfv\'{e}n waves depend strongly on
the value of the plasma $\beta$, which is defined here as the
ratio of proton plasma pressure to the magnetic pressure
(see Gary \& Borovsky 2004, 2008).
For the high-latitude heliocentric distances considered here,
we used a very simple monopolar scaling for the magnetic field
strength,
\begin{equation}
  B_{0}(r) \, \approx \, 2.5 \times 10^{-5}
  \left( \frac{r}{1 \,\, \mbox{AU}} \right)^{-2}
  \, \mbox{G} 
\end{equation}
and $\beta = 8\pi n_{p} k_{\rm B} T_{p} / B_{0}^2$.
Using the empirically constrained densities and temperatures
discussed above, the value of $\beta$ increases monotonically
from 1.1 (at 0.3 AU) to about 37 (at 5 AU).
Because calculating the dispersive properties of the waves on a
fine two-dimensional grid in wavenumber is computationally expensive,
we created only 7 grids having $\beta = 1,$ 1.8, 3.4, 6.4, 12, 22,
and 40.
We computed full runs of $Q_p$ and $Q_e$ versus heliocentric
distance using each of the 7 grids, and then we interpolated
between these results (using the empirical plasma $\beta$ at
each distance) in order to determine the appropriate solutions.

Figure 5 shows the results of this procedure for the slab, 2D,
and isotropic spectra, and we compare the proton heating fractions
$Q_{p} / (Q_{p} + Q_{e})$ to the standard empirical curve from
Figures 3 and 4.
The curve denoting the slab spectrum is relatively simple to
understand.
In a collisionless plasma, parallel-propagating Alfv\'{e}n waves
dissipate primarily by cyclotron resonance with positive ions
(see Hollweg \& Isenberg 2002).
For the values of $\beta$ used in the warm-plasma dispersion code,
the largest wave frequency for parallel-propagating waves was
only about 0.3 times the local proton gyrofrequency.
Thus, since the model did not contain any other ions, only the
protons felt the cyclotron resonance and were responsible for
almost all of the dissipation.
There was some extremely weak Landau damping in the slab model
that contributed about 0.1\% of the total heating to the electrons.

\begin{figure}
\epsscale{1.13}
\plotone{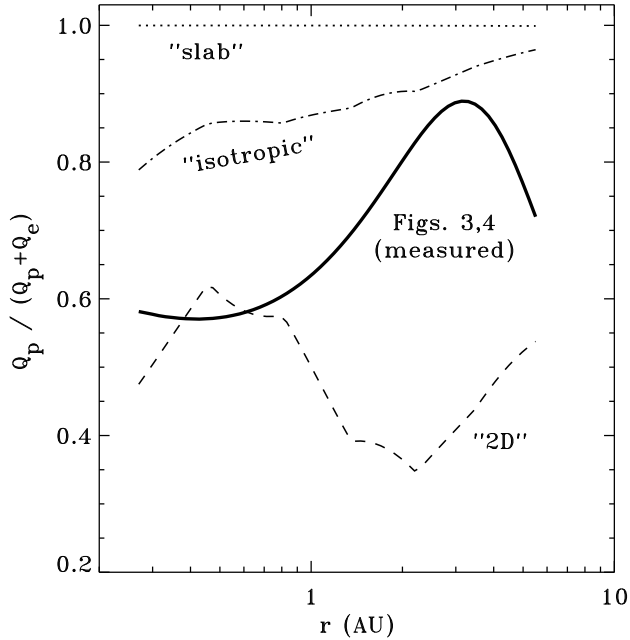}
\caption{Comparison of the standard empirical proton-to-total
heating ratio ({\em{solid line}}) to theoretical ratios computed
from the linear damping of Alfv\'{e}n wave spectra:
parallel-propagating ``slab'' modes only ({\em{dotted line}}),
perpendicular ``2D'' modes only ({\em{dashed line}}), and
an isotropic spectrum ({\em{dot-dashed line}}).}
\end{figure}

The curve denoting the 2D spectrum (i.e., fluctuations perpendicular
to the background magnetic field) always shows a roughly equal
partitioning of energy between protons and electrons.
However, the partition fraction oscillates slightly up and down
several times between 0.3 and 5 AU.
These oscillations do not seem to be numerical artifacts, and
their origin is not entirely clear.
For the range of plasma $\beta$ found in these models, there
is never a single physical process that dominates the
collisionless damping of highly oblique Alfv\'{e}n waves.
The mechanisms of electron Landau damping, proton Landau damping,
and proton transit-time damping are all of comparable magnitude
(e.g., Gary \& Borovsky 2004, 2008).
Thus, we refrain from attributing the mild variations around the
mean equipartitioning ($Q_{p} \approx Q_{e}$) to any specific
physical process.
We propose to study this complex interplay of collisionless
dissipation mechanisms in future work.
Finally, the curve denoting an isotropic fluctuation spectrum
tends gives an intermediate amount of heat partitioning between
the slab and 2D cases, as one would expect.

By comparing the theoretical and empirical curves in Figure 5,
we can make some preliminary suggestions about how the anisotropy
of the actual turbulent power spectrum (or at least its
large-wavenumber dissipation range) varies between 0.3 and 5 AU.
An interpretation based on Figure 5 is that the fluctuations are
close to 2D at the smallest heliocentric distances, and that they
evolve to be more isotropic as distance is increased.
This makes heuristic sense from at least one limited viewpoint:
i.e., that as the plasma $\beta$ increases and the effect of
the magnetic field becomes less important to the plasma, the
turbulence may want to behave in an increasingly hydrodynamic
(i.e., isotropic) fashion.
Note, however, that this conclusion is predicated on a
wide range of assumptions about the models.
Most importantly, it depends on the turbulent fluctuations
being purely in the Alfv\'{e}n mode, and not having, say,
fast-mode/whistler type dispersive properties.
This conclusion also depends on the turbulent fluctuations
damping like linear waves, despite our knowledge that nonlinear
features such as current sheets may play an important role
(Dmitruk et al.\  2004).

We should also note that if the fit to the electron heat conduction
(eq.~[\ref{eq:qefit}]) is incorrect, and instead the collisionless
expression of Hollweg (1974, 1976) is taken to be valid, then
the above conclusions may be {\em reversed.}
A comparison between the heating partition curve that was computed
for $\alpha_{e} = 1.05$ (from Fig.\ 3{\em{b}}) with the theoretical
results in Figure 5 would imply that the turbulence evolves from a
slab or isotropic spectrum at 0.3 AU to a more 2D spectrum at
larger distances.
This serves to emphasize how much our conclusions about proton and
electron heating depend on the accurate knowledge of the radial
dependence of $q_{\parallel, e}$ in the fast solar wind.

There has been a great deal of work done to measure the
anisotropy of MHD turbulence in the solar wind.
Ultimately, the empirical data for {\em both} the particle
heating and the fluctuation spectra should be combined to
provide the tightest possible constraints on theoretical
models.
At 1 AU, the dominant component of the turbulence does seem
to be perpendicular or 2D (e.g., Bieber et al.\  1996;
MacBride et al.\  2008).
However, when analyzing the radial dependence of the anisotropy,
there have been differing answers to some of the key questions.
Some studies that compared the power between the minimum and
maximum variance directions concluded that the turbulence becomes
more isotropic as distance increases (Klein et al.\  1991;
Horbury et al.\  1995).
Other studies found that the perpendicular component becomes
more dominant at larger distances (Bavassano et al.\  1982;
Neugebauer 2004).
Such opposite conclusions may be the result of mixing together
regions that contain different amounts of fast and slow wind
(see, e.g., Dasso et al.\  2005) or using different ranges of
frequency.
Also, the variance anisotropy does not necessarily vary in the
same way as the spectral (wavenumber) anisotropy.
These results may also depend on subtleties of the analysis
technique, such as how the minimum-variance direction is
interpreted (Smith et al.\  2006b; Tessein et al.\  2009) or how
effects such as intermittency are treated (Bruno et al.\  1999).

\section{Conclusions and Future Prospects}

The goal of this paper has been to compute empirical estimates
for the rates of proton and electron heating in the fast solar
wind between 0.29 and 5.4 AU.
A key new aspect of this work has been the incorporation of the
measured electron heat conduction flux, which dominates the
internal energy balance of electrons in much of the heliosphere.
We conclude that the protons receive about 60\% of the total
plasma heating in the inner heliosphere, and that this fraction
increases to approximately 80\% by the orbit of Jupiter.
These results are confirmed by the independent analyses of
Leamon et al.\  (1999) and Breech et al.\  (2009), who found that
a similar proton-to-total heating fraction of $\sim$60\% is
consistent with theoretical models of MHD turbulent dissipation.

Various uncertainties in the measurements affected our results.
For example, the rates of proton and electron heating ($Q_p$
and $Q_e$) were calculated from mean radial trends in the data
(as exemplified by the least-squares fits given in {\S}~2), and
they do not describe the substantial measured {\em spread} in
the data about the mean values.
Also, the uncertainty in $Q_e$ increases with heliocentric
distance because the electron heat conduction flux becomes more
important---in a relative sense---as distance increases.
Small uncertainties in $q_{\parallel, e}$ or its radial slope
thus have a larger impact on the derived heating rates at
distances greater than $\sim$2 AU.
In any case, the general techniques presented in this paper
are applicable to any future measurements that would improve on
our current knowledge of the solar wind plasma properties.

It is interesting to compare the interplanetary proton and
electron heating rates with those inferred for the solar corona.
The observational consensus since the late 1990s has been that
coronal holes undergo preferential proton heating in a similar
sense as high-speed wind streams in the heliosphere (see, e.g.,
Kohl et al.\  1997, 1998, 2006; Wilhelm et al.\  2007).
Recently, Figure 6 of Landi \& Cranmer (2009) summarized a range
of measurements of $T_p$ and $T_e$ in polar coronal holes.
Although one generally finds $T_{p} > T_{e}$, the measurements
of $T_p$ and $T_e$ do not yet fully overlap with one another in
heliocentric distance.
Improved measurements are needed in order to better constrain the
proton and electron heating rates in the corona.
However, given that the corona and the heliosphere are so different
in density, Alfv\'{e}n speed, and plasma $\beta$, it is somewhat
surprising that the proton-electron heat partitioning in these
regions may be so similar in character.

An improved understanding of the so-called {\em two-fluid} nature
of solar wind plasma is an important ingredient in producing
better quantitative predictions of the heliospheric properties
relevant to space weather.
Most advanced global-modeling efforts currently include only a
one-fluid treatment of the internal energy equation (e.g., 
Riley et al.\  2001, 2006; Roussev et al.\  2003;
T\'{o}th et al.\  2005; Usmanov \& Goldstein 2006;
Feng et al.\  2007).
It is often assumed that in the highest density (i.e., most
strongly collisional) regions of the heliosphere, the departures
from thermal equilibrium are unimportant.
However, there are several kinds of large-scale effects
that depend on how heat is deposited into protons, electrons,
and possibly heavy ions as well.
For example, if all of the coronal heating goes into electrons,
there can be substantially more downward conduction than in the
proton-heated case, which would affect the location of the coronal
temperature maximum (Hansteen \& Leer 1995) and the dynamical
stability of helmet streamers (Endeve et al.\  2004).
Thus, the eventual inclusion of differences between proton heating
and electron heating in global models may be a key to improving
their physical realism and predictive accuracy.

\acknowledgments

The authors would like to thank Adriaan van Ballegooijen
and Eliot Quataert for valuable discussions.
SRC's work was supported by the National Aeronautics and Space
Administration (NASA) under grants {NNG\-04\-GE77G,}
{NNX\-06\-AG95G,} and {NNX\-09\-AB27G}
to the Smithsonian Astrophysical Observatory.
WHM's research was supported by NSF ATM 0752135 (SHINE) and 
NASA {NNX\-08\-AI47G} (Heliophysics Theory Program).
BAB's research was supported in part by an appointment to the
NASA Postdoctoral Program at Goddard Space Flight Center,
administered by Oak Ridge Associated Universities through
a contract with NASA.
JCK's research was supported in part by NASA grant NNX08AW07G.


\begin{thebibliography}{}

\bibitem[]{AF91}
Arya, S. \& Freeman, J. W. 1991, \jgr, 96, 14183

\bibitem[]{Ba92}
Bame, S. J., McComas, D. J., Barraclough, B. L., Phillips, J. L.,
Sofaly, K. J., Chavez, J. C., Goldstein, B. E., \& Sakurai, R. K.
1992, \aaps, 92, 237

\bibitem[]{BS82}
Barakat, A. R., \& Schunk, R. W. 1982, Plasma Phys., 24, 389

\bibitem[]{Bv82}
Bavassano, B., Dobrowolny, M., Fanfoni, G., Mariani, F., \& Ness,
N. F. 1982, \solphys, 78, 373

\bibitem[]{Bv00}
Bavassano, B., Pietropaolo, E., \& Bruno, R. 2000, \jgr, 105, 15959

\bibitem[]{Bb96}
Bieber, J. W., Wanner, W., \& Matthaeus, W. H. 1996, \jgr, 101, 2511

\bibitem[]{Bo05}
Boldyrev, S. 2005, \apj, 626, L37

\bibitem[]{Br65}
Braginskii, S. I. 1965, Rev.\  Plasma Phys., 1, 205

\bibitem[]{Br09}
Breech, B., Matthaeus, W. H., Cranmer, S. R., Kasper, J., \&
Oughton, S. 2009, \jgr, in press

\bibitem[]{Br08a}
Breech, B., Matthaeus, W. H., Minnie, J., Bieber, J. W., Oughton, S.,
Smith, C. W., \& Isenberg, P. A.  2008, \jgr, 113, A08105

\bibitem[]{Bu99}
Bruno, R., Bavassano, B., Pietropaolo, E., Carbone, V., \& Veltri, P.
1999, \grl, 26, 3185

\bibitem[]{Bu86}
Bruno, R., \& Dobrowolny, M. 1986, Ann.\  Geophys., 4, 17

\bibitem[]{CV00}
Cho, J., \& Vishniac, E. T. 2000, \apj, 539, 273

\bibitem[]{Co68}
Coleman, P. J., Jr. 1968, \apj, 153, 371

\bibitem[]{Cr02}
Cranmer, S. R. 2002, \ssr, 101, 229

\bibitem[]{CFK99}
Cranmer, S. R., Field, G. B., \& Kohl, J. L. 1999, \apj, 518, 937

\bibitem[]{CvB03}
Cranmer, S. R., \& van Ballegooijen, A. A. 2003, \apj, 594, 573

\bibitem[]{CvB05}
Cranmer, S. R., \& van Ballegooijen, A. A. 2005, \apjs, 156, 265

\bibitem[]{CH70}
Cuperman, S., \& Harten, A. 1970, \apj, 162, 315

\bibitem[]{Ds05}
Dasso, S., Milano, L. J., Matthaeus, W. H., \& Smith, C. W. 2005,
\apjl, 635, L181

\bibitem[]{Dm02}
Dmitruk, P., Matthaeus, W. H., Milano, L. J., Oughton, S., Zank,
G. P., \& Mullan, D. J. 2002, \apj, 575, 571

\bibitem[]{Dm04}
Dmitruk, P., Matthaeus, W. H., \& Seenu, N. 2004, \apj, 617, 667

\bibitem[]{Dm01}
Dmitruk, P., Milano, L. J., \& Matthaeus, W. H. 2001, \apj, 548, 482

\bibitem[]{Db80}
Dobrowolny, M., Mangeney, A., \& Veltri, P. 1980, \prl, 45, 144

\bibitem[]{Du83}
Dum, C. T. 1983, in Solar Wind Five, ed. M. Neugebauer, NASA CP-2280,
369

\bibitem[]{Eb09}
Ebert, R. W., McComas, D. J., Elliott, H. A., Forsyth, R. J., \&
Gosling, J. T. 2009, \jgr, 114, A01109

\bibitem[]{Ei79}
Eichler, D. 1979, \apj, 229, 413

\bibitem[]{En04}
Endeve, E., Holzer, T. E., \& Leer, E. 2004, \apj, 603, 307

\bibitem[]{Ey81}
Eyni, M., \& Steinitz, R. 1981, \apj, 243, 279

\bibitem[]{Fx07}
Feng, X., Zhou, Y., \& Wu, S. T. 2007, \apj, 655, 1110

\bibitem[]{Fr88}
Freeman, J. W. 1988, \grl, 15, 88

\bibitem[]{Fy77}
Fyfe, D., Joyce, G., \& Montgomery, D. 1977, J.\  Plasma Phys., 17, 317

\bibitem[]{GB04}
Gary, S. P., \& Borovsky, J. E. 2004, \jgr, 109, A06105

\bibitem[]{GB08}
Gary, S. P., \& Borovsky, J. E. 2008, \jgr, 113, A12104

\bibitem[]{GS95}
Goldreich, P., \& Sridhar, S. 1995, \apj, 438, 763

\bibitem[]{Gm96}
Goldstein, B. E., et al. 1996, \aap, 316, 296

\bibitem[]{Ge95}
Goldstein, M. L., Roberts, D. A., \& Matthaeus, W. H. 1995, \araa,
33, 283

\bibitem[]{Gr90}
Grappin, R., Mangeney, A., \& Marsch, E. 1990, \jgr, 95, 8197

\bibitem[]{HL95}
Hansteen, V. H., \& Leer, E. 1995, \jgr, 100, 21577

\bibitem[]{He06}
Hellinger, P., Tr\'{a}vn\'{\i}\v{c}ek, P., Kasper, J. C., \&
Lazarus, A. J. 2006, \grl, 33, L09101

\bibitem[]{Hg84}
Higdon, J. C. 1984, \apj, 285, 109

\bibitem[]{Ho74}
Hollweg, J. V. 1974, \jgr, 79, 3845

\bibitem[]{Ho76}
Hollweg, J. V. 1976, \jgr, 81, 1649

\bibitem[]{HI02}
Hollweg, J. V., \& Isenberg, P. A. 2002, \jgr, 107 (A7), 1147

\bibitem[]{Hb95}
Horbury, T. S., Balogh, A., Forsyth, R .J., \& Smith, E. J. 1995,
\grl, 22, 3405

\bibitem[]{Hb08}
Horbury, T. S., Forman, M., \& Oughton, S. 2008, \prl, 101, 175005

\bibitem[]{Hs95}
Hossain, M., Gray, P. C., Pontius, D. H., Jr., Matthaeus, W. H.,
\& Oughton, S. 1995, Phys.\  Fluids, 7, 2886

\bibitem[]{Ir64}
Iroshnikov, P. S. 1964, \sovast, 7, 566

\bibitem[]{I84}
Isenberg, P. A. 1984, \jgr, 89, 6613

\bibitem[]{Is98}
Issautier, K., Meyer-Vernet, N., Moncuquet, M., \& Hoang, S. 1998,
\jgr, 103, 1969

\bibitem[]{Ka02}
Kasper, J. C., Lazarus, A. J., \& Gary, S. P. 2002, \grl, 29, 1839

\bibitem[]{Ka08}
Kasper, J. C., Lazarus, A. J., \& Gary, S. P. 2008, \prl, 101, 261103

\bibitem[]{Kg00}
Kellogg, P. J. 2000, \apj, 528, 480

\bibitem[]{Kl91}
Klein, L. W., Roberts, D. A., \& Goldstein, M. L. 1991, \jgr, 96, 3779

\bibitem[]{K1st}
Kohl, J. L., et al. 1997, \solphys, 175, 613

\bibitem[]{Klett}
Kohl, J. L., et al. 1998, \apjl, 501, L127

\bibitem[]{K06}
Kohl, J. L., Noci, G., Cranmer, S. R., \& Raymond, J. C. 2006,
\aapr, 13, 31

\bibitem[]{K41}
Kolmogorov, A. N. 1941, Dokl.\  Akad.\  Nauk SSSR, 30, 301

\bibitem[]{Kr65}
Kraichnan, R. H. 1965, Phys.\  Fluids, 8, 1385

\bibitem[]{LC09}
Landi, E., \& Cranmer, S. R. 2009, \apj, 691, 794

\bibitem[]{Le99}
Leamon, R. J., Smith, C. W., Ness, N. F., \& Wong, H. K. 1999,
\jgr, 104, 22331

\bibitem[]{LHF}
Leer, E., Holzer, T. E., \& {Fl\aa}, T. 1982, \ssr, 33, 161

\bibitem[]{LF86}
Lopez, R. E., \& Freeman, J. W. 1986, \jgr, 91, 1701

\bibitem[]{Mc08}
MacBride, B. T., Smith, C. W., \& Forman, M. A. 2008, \apj, 679, 1644

\bibitem[]{Mr08}
Marino, R., Sorriso-Valvo, L., Carbone, V., Noullez, A., Bruno, R.,
\& Bavassano, B. 2008, \apjl, 677, L71

\bibitem[]{Ma99}
Marsch, E. 1999, \ssr, 87, 1

\bibitem[]{Ma82}
Marsch, E., M\"{u}hlh\"{a}user, K.-H., Schwenn, R., Rosenbauer, H.,
Pilipp, W., \& Neubauer, F. M. 1982, \jgr, 87, 52

\bibitem[]{MT01}
Marsch, E., \& Tu, C.-Y. 2001, \jgr, 106, 227

\bibitem[]{Mn07}
Matteini, L., Landi, S., Hellinger, P., Pantellini, F., Maksimovic, M.,
Velli, M., Goldstein, B. E., \& Marsch, E. 2007, \grl, 34, L20105

\bibitem[]{Ma05}
Matthaeus, W. H., Dasso, S., Weygand, J. M., Milano, L. J.,
Smith, C. W., \& Kivelson, M. G. 2005, \prl, 95, 231101

\bibitem[]{Ma03}
Matthaeus, W. H., Dmitruk, P., Oughton, S., \& Mullan, D. 2003, in
Solar Wind Ten, AIP Conf.\  Ser.\  679, ed. M. Velli \& R. Bruno
(Woodbury, NY: AIP), 427

\bibitem[]{Mt99a}
Matthaeus, W. H., Zank, G. P., Oughton, S., Mullan, D. J., \&
Dmitruk, P. 1999a, \apjl, 523, L93

\bibitem[]{Mt99b}
Matthaeus, W. H., Zank, G. P., Smith, C. W., \& Oughton, S. 1999b,
\prl, 82, 3444

\bibitem[]{MiLa}
Miller, J. A., LaRosa, T. N., \& Moore, R. L.  1996, \apj, 461, 445

\bibitem[]{Nk01}
Nakayama, K. 2001, \apj, 556, 1027

\bibitem[]{Ng82}
Neugebauer, M. 1982, \ssr, 33, 127

\bibitem[]{Ng04}
Neugebauer, M. 2004, \jgr, 109, A02101

\bibitem[]{OL96}
Olsen, E. L., \& Leer, E. 1996, \jgr, 101, 15591

\bibitem[]{Ou01}
Oughton, S., Matthaeus, W. H., Dmitruk, P., Milano, L. J., Zank,
  G. P., \& Mullan, D. J. 2001, \apj, 551, 565

\bibitem[]{P63}
Parker, E. N. 1963, Interplanetary Dynamical Processes (New York:
Interscience)

\bibitem[]{Pk73}
Perkins, F. 1973, \apj, 179, 637

\bibitem[]{Ph95}
Phillips, J. L., Bame, S. J., Gary, S. P., Gosling, J. T., Scime,
E. E., \& Forsyth, R. J. 1995, \ssr, 72, 109

\bibitem[]{Ph90}
Phillips, J. L., \& Gosling, J. T. 1990, \jgr, 95, 4217

\bibitem[]{Pi90}
Pilipp, W. G., Miggenrieder, H., M\"{u}hlh\"{a}user, K.-H.,
Rosenbauer, H., \& Schwenn, R. 1990, \jgr, 95, 6305

\bibitem[]{Q98}
Quataert, E. 1998, \apj, 500, 978

\bibitem[]{QG99}
Quataert, E., \& Gruzinov, A. 1999, \apj, 520, 248

\bibitem[]{Ry01}
Riley, P., Linker, J. A., \& Miki\'{c}, Z. 2001, \jgr, 106, 15889

\bibitem[]{Ry06}
Riley, P., Linker, J. A., Miki\'{c}, Z., Lionello, R., Ledvina, S. A.,
\& Luhmann, J. G. 2006, \apj, 653, 1510

\bibitem[]{Ro03}
Roussev, I. I., Gombosi, T. I., Sokolov, I. V., Velli, M.,
Manchester, W., IV, DeZeeuw, D. L., Liewer, P., T\'{o}th, G.,
\& Luhmann, J. 2003, \apjl, 595, L57

\bibitem[]{Se03}
Salem, C., Hubert, D., Lacombe, C., Bale, S. D., Mangeney, A.,
Larson, D. E., \& Lin, R. P. 2003, \apj, 585, 1147

\bibitem[]{Sa95}
Sandb{\ae}k, {\O}., \& Leer, E. 1995, \apj, 454, 486

\bibitem[]{SM90}
Schwenn, R., \& Marsch, E., eds. 1990, Physics of the Inner Heliosphere
(Heidelberg: Springer-Verlag)

\bibitem[]{Sc99}
Scime, E. E., Badeau, A. E., Jr., \& Littleton, J. E. 1999, \grl,
26, 2129

\bibitem[]{Sc94}
Scime, E. E., Bame, S. J., Feldman, W. C., Gary, S. P., Phillips,
J. L., \& Balogh, A. 1994, \jgr, 99, 23401

\bibitem[]{SO79}
Scudder, J. D., \& Olbert, S. 1979, \jgr, 84, 6603

\bibitem[]{SS80}
Sittler, E. C., Jr., \& Scudder, J. D. 1980, \jgr, 85, 5131

\bibitem[]{Sm01}
Smith, C. W., Matthaeus, W. H., Zank, G. P., Ness, N. F., Oughton, S.,
\& Richardson, J. D. 2001, \jgr, 106, 8253

\bibitem[]{Sm06a}
Smith, C. W., Isenberg, P. A., Matthaeus, W. H., \& Richardson, J. D.
2006a, \apj, 638, 508

\bibitem[]{Sm06b}
Smith, C. W., Vasquez, B. J., \& Hamilton, K. 2006b, \jgr, 111, A09111

\bibitem[]{Sp91}
Spangler, S. R. 1991, \apj, 376, 540

\bibitem[]{Sp62}
Spitzer, L., Jr. 1962, Physics of Fully Ionized Gases, 2nd ed.
(New York: Wiley)

\bibitem[]{Sp53}
Spitzer, L., Jr., \& H\"{a}rm, R. 1953, Phys.\  Rev., 89, 977

\bibitem[]{Sw09}
Stawarz, J. E., Smith, C. W., Vasquez, B. J., Forman, M. A., \&
MacBride, B. T. 2009, \apj, 697, 1119

\bibitem[]{Stix}
Stix, T. H. 1992, Waves in Plasmas (New York: AIP)

\bibitem[]{Te09}
Tessein, J. A., Smith, C. W., MacBride, B. T., Matthaeus, W. H.,
  Forman, M. A., \& Borovsky, J. E. 2009, \apj, 692, 684

\bibitem[]{Tg05}
T\'{o}th, G., et al. 2005, \jgr, 110, A12226

\bibitem[]{Tt95}
Totten, T. L., Freeman, J. W., \& Arya, S. 1995, \jgr, 100, 13

\bibitem[]{Tu88}
Tu, C.-Y. 1988, \jgr, 93, 7

\bibitem[]{TM95}
Tu, C.-Y., \& Marsch, E. 1995, \ssr, 73, 1

\bibitem[]{Us06}
Usmanov, A. V., \& Goldstein, M. L. 2006, \jgr, 111, A07101

\bibitem[]{Vb07}
Vasquez, B. J., Smith, C. W., Hamilton, K., MacBride, B. T., \&
Leamon, R. J. 2007, \jgr, 112, A07101

\bibitem[]{Vr95}
Verma, M. K., Roberts, D. A., \& Goldstein, M. L. 1995, \jgr,
100, 19839

\bibitem[]{Wh07}
Wilhelm, K., Marsch, E., Dwivedi, B. N., \& Feldman, U. 2007,
\ssr, 133, 103

\bibitem[]{Wi95}
Williams, L. L. 1995, \apj, 453, 953

\bibitem[]{Za96}
Zank, G. P., Matthaeus, W. H., \& Smith, C. W. 1996, \jgr, 101, 17093

\bibitem[]{ZM90}
Zhou, Y., \& Matthaeus, W. H. 1990, \jgr, 95, 10291

\end{thebibliography}
\end{document}